\documentclass[jcp,twocolumn,preprintnumbers,nourl,superscriptaddress]{revtex4-1}

\usepackage{amsmath,amssymb}
\usepackage{graphicx,psfrag}
\usepackage{dcolumn}
\usepackage{bm,tabularx,ragged2e,booktabs,caption}
\usepackage{ulem}
\usepackage{amsmath}
\usepackage{setspace}
\usepackage{leftidx}
\usepackage[USenglish]{babel}

\usepackage{amssymb,amsfonts,amsmath}
\usepackage{bm}

\usepackage[T1]{fontenc}
\usepackage[latin1]{inputenc}

\numberwithin{equation}{section}

\numberwithin{equation}{section}

\begin{document}

\title{More accurate and efficient bath spectral densities from super-resolution}

\author{Thomas Markovich}
\affiliation{Department of Chemistry and Chemical Biology, Harvard
University, Cambridge, MA 02138, USA}
\author{Samuel M. Blau}
\affiliation{Department of Chemistry and Chemical Biology, Harvard
University, Cambridge, MA 02138, USA}
\author{John Parkhill}
\affiliation{Department of Chemistry, The University of Notre Dame, South Bend, IN 46556, USA}
\author{Christoph Kreisbeck}
\affiliation{Department of Chemistry and Chemical Biology, Harvard
University, Cambridge, MA 02138, USA}
\author{Jacob N. Sanders}
\affiliation{Department of Chemistry and Chemical Biology, Harvard
University, Cambridge, MA 02138, USA}
\author{Xavier Andrade}
\affiliation{Department of Chemistry and Chemical Biology, Harvard
University, Cambridge, MA 02138, USA}
\author{Al\'{a}n Aspuru-Guzik}
\affiliation{Department of Chemistry and Chemical Biology, Harvard
University, Cambridge, MA 02138, USA}
\email{aspuru@chemistry.harvard.edu}

\begin{abstract}
	Quantum transport and other phenomena are typically modeled by coupling the system of interest to an environment, or bath, held at thermal equilibrium.  Realistic bath models are at least as challenging to construct as models for the quantum systems themselves, since they must incorporate many degrees of freedom that interact with the system on a wide range of timescales.  Owing to computational limitations, the environment is often modeled with simple functional forms, with a few parameters fit to experiment to yield semi-quantitative results.  Growing computational resources have enabled the construction of more realistic bath models from molecular dynamics (MD) simulations.  In this paper, we develop a numerical technique to construct these atomistic bath models with better accuracy and decreased cost.  We apply a novel signal processing technique, known as super-resolution, combined with a dictionary of physically-motivated bath modes to derive spectral densities from MD simulations.  Our approach reduces the required simulation time and provides a more accurate spectral density than can be obtained via standard Fourier transform methods.  Moreover, the spectral density is provided as a convenient closed-form expression which yields an analytic time-dependent bath kernel. Exciton dynamics of the Fenna-Matthews-Olsen light-harvesting complex are simulated with a second order time-convolutionless master equation, and spectral densities constructed via super-resolution are shown to reproduce the dynamics using only a quarter of the amount of MD data. 

\end{abstract}


\maketitle


\section{Introduction}
\indent  Irreversible processes such as solvation, energy transfer, and chemical binding have received renewed interest in recent years.  Because these processes involve large systems with many degrees of freedom, the typical approach to studying these processes is the open quantum systems formalism, in which the degrees of freedom are partitioned into a system of interest and a bath held at thermal equilibrium~\cite{Breuer:2007wp, Shenvi:2008tr}.  It is commonly assumed that the system only couples weakly to the bath, making the precise nature of the bath a secondary concern in the physical theory.  For example, in studying the energy transfer dynamics in a system of chromophores embedded in a protein framework, each chromophore is individually coupled to many thousands of atoms in the protein, but the system-bath formalism dramatically simplifies all of these couplings in order to make the dynamics tractable~\cite{2013JChPh.138k4102B,Berkelbach:2013vw, Singh:2011fl}. Renewed interest in the strong and intermediate coupling region, relevant for energy transfer in the exciton dynamics of light-harvesting complexes, has lead to various studies~\cite{Jang:2007cz, Mohseni:2008gp, Plenio:2008ff, Rebentrost:2009hu, Cao:2009cc, Ishizaki:2009tt, Ishizaki:2011cx, Sarovar:2010hs, Abramavicius:2010et,Wu:2010bg,Moix:2011gi,Kreisbeck:2011dh,Skochdopole:2011gh,Ritschel:2011ht,Rebentrost:2011hc,Singh:2011fl,Pachon:2012fm,Vlaming:2012hv,Caruso:2012gf,Zhu:2011ea,Roden:2009gr,Olbrich:2010ce,Olbrich:2011hc,Hein:2012vn} on the precise influence of the bath on the higher systems. Higher order phonon processes, non-Markovian effects and structures in the exciton-phonon coupling change the energy transfer~\cite{doi:10.1021/jp304649c,Chin:2013uh, Kreisbeck:2012ui}. Thus, details in the bath are relevant and need to be taken into account in realistic simulations. Accordingly, our goal in this paper is to apply a recent signal-processing technique known as super-resolution to obtain realistic atomistic models of environments containing thousands of atoms at feasible computational expense.  With these atomistic bath models in hand, one can begin to evaluate the importance of a realistic bath model in a physical theory.

In the approach to open quantum systems employed in this work, we model the bath by an ensemble of noninteracting harmonic oscillators. The central mathematical object of such a model is the {\it spectral density}, $J(\omega)$, which gives the frequency-dependent strength of system-bath coupling. The spectral density can be understood as the density of bath oscillator states at each frequency.  Owing to computational limitations, most studies of open quantum systems assume an extremely simple functional form for the spectral density, such as a single broad peak covering all relevant excitonic transitions of the system.  With the goal of providing more physically accurate bath models and dynamics, Valleau {\it et al.} has previously obtained atomistic spectral densities for the Fenna-Matthews-Olson (FMO) complex from combined Molecular Dynamics (MD)~\cite{Valleau:2012ig, Tuckerman:2010wy} and Time-Dependent Density Functional Theory (TDDFT)~\cite{Runge:1984us} simulations.  However, the difficulty of this more realistic approach is the high computational cost of running expensive TDDFT calculations at every step in an MD simulation.  In order to obtain a spectral density of sufficient resolution, the MD-TDDFT simulation must be run for over 40 picoseconds (ps)~\cite{Mallat:2008vn}, which may become computationally intractable for larger systems.

To make progress, we first observe that a typical vibrational bath is not an arbitrary function but rather a relatively sparse collection of damped harmonic oscillators.  Sparsity enables us to apply a novel numerical technique known as super-resolution in order to reconstruct the spectral density from much shorter MD-TDDFT simulations. Super-resolution has been applied to a broad range of scientific problems, including image~\cite{Freeman:2002va} and video compression~\cite{Patti:1997vq}, image denoising~\cite{Elad:1997un}, astronomy~\cite{2005A&A...436..373P}, microscopy~\cite{Mccutchen:1967uf}, and medical imaging~\cite{5193030}. To our knowledge, this paper is the first application of super-resolution to quantum dynamics. Super-resolution provides a provably convergent algorithm for the reconstruction of signals from limited time-domain measurements using a total variation minimization procedure. Super-resolution is related to compressed sensing~\cite{5272200,Donoho:ci,5288845,4770164,4959603,MRM:MRM21391,6426647,Tuma:2009gb,Shabani:2009de,5419072,4472247,Coulter:2010wx}. Compressed sensing is a technique designed to recover sparse signals from randomly-sampled data by minimizing the $\mathcal{L}_1$ norm of an underdetermined system of linear equations. Compressed sensing works by finding the sparsest signals consistent with the underdetermined system of equations. This usually involves an optimization problem. Despite its success in many applications, the $\mathcal{L}_1$-norm minimization of compressed sensing can result in spurious signals as it emphasizes the sparsity of the solution only. Super-resolution is a numerical method that shares the spirit of compressed sensing. The difference between superresolution and compressed sensing stems from both the choice of objective function and sampling technique. It was developed to recover sparse signals from nonrandomly undersampled data. By minimizing the $\mathcal{L}_1$-norm of the gradient of the function in addition to the $\mathcal{L}_1$ norm of the function itself, super-resolution allows for smoother solutions to the sampling problem~\cite{CPA:CPA21455, BioucasDias:2007tp, 4378902}.


Because of the ample experimental and theoretical data to compare against~\cite{Valleau:2012ig,Moix:2011gi,Vulto:1998bu,Olbrich:2011hc,Adolphs:2006ey,Mohseni:2011wg,Chen:2013wz,YuenZhou:2012vf,Ritschel:2011ht,Shim2012649,Pachon:2012fm,Vlaming:2012hv,Plenio:2008ff,Singh:2011fl,Kolli:2011vy,Rebentrost:2009hu,Skochdopole:2011gh,Kreisbeck:2011dh,Kreisbeck:2012ui,Zhu:2011ea,Cao:2009cc,Sarovar:2010hs,Ishizaki:2009tt,Jang:2011vc,Ishizaki:2009uh}, we apply super-resolution to the FMO light-harvesting complex of {\it C. tepidium} but emphasize that this technique is broadly applicable. While this paper focuses on a vibrational bath which perturbs the energies of molecular electronic states, the techniques we introduce are generic for any model of a bath which is based on time-correlation functions.

\section{Super-resolution of Spectral Densities}

In this section, we briefly review the procedure for simulating the dynamics of open quantum systems and computing spectral densities from combined MD-TDDFT simulations. We then apply the theory of super-resolution to accelerate and improve the accuracy of these computations.  Computing spectral densities from atomistic calculations, rather than from semi-empirical functional forms, enables the inclusion of molecular vibrations and other physical effects (such as solvation effects) to produce a more realistic bath model~\cite{Valleau:2012ig}.  Super-resolution, in turn, brings the construction of these atomistic bath models into the realm of computational feasibility.

Armed with our more realistic bath model, we will employ a
second-order time-convolution\-less master equation (TCL-2) to
simulate the dynamics of FMO monomer, allowing us to evaluate the
physical impact of different approximations to the spectral density.
TCL-2 includes non-Markovian effects up to second order in the
system-bath coupling. By comparing TCL-2 with exact methods like the
hierarchical equations of motion (HEOM)~\cite{JPSJ.58.101} we show that
most of the relevant effects of the structured spectral density of the
FMO complex are captured by TCL-2. Here we use TCL-2, since it is
numerically more treatable than HEOM, in particular for structured
spectral densities where HEOM becomes cumbersome and requires a high
performance GPU
implementation~\cite{Kreisbeck:2011dh,Kreisbeck:2012ui, 16106}. We
employ the equation of
motion~\cite{Breuer:2007wp,Rebentrost:2011vh,Rebentrost:2009hu,Shim2012649,Mohseni:2008gp,Zhu:2011ea,
  2006JChPh.125j4906P,2011PhRvB..83k5416T,Ahn:1994ww,HeinzPeter:2000ul,Shabani:2005wa,Smirne:2010te}:
\begin{align}
\label{eq:tcl}
& \frac{d \rho_{I}(t)}{dt} = -\frac{i}{\hbar} [ H_{I},\rho_{I} ] \\
				& - \frac{1}{\hbar^2} \displaystyle\sum_{n}  \int_0^{t} \!\!\! \mathrm{d}\tau \, D_{n}(t-\tau)[H_{In}(t),[H_{In}(\tau), \rho_{I}(t)]]  \nonumber\\
	&	D_{n}(t) = \\ 
&	\int_{0}^{\infty} \!\!\! \mathrm{d}\omega \, \,   J_{n}(\omega) \left[\coth \! \left( \! \frac{\hbar \, \omega \beta \,}{2} \! \right) \cos(\omega \, t)  - i \sin(\omega \, t) \right] 	 \nonumber 		
\end{align}
where $H$ is the system Hamiltonian, $\rho$ is the system density matrix, $D$ is our bath kernel, the subscript $I$ indicates that we are in the interaction picture, the summation runs over all sites, and $J(\omega)$ is the spectral density computed via super-resolution~\cite{Kolli:2011vy, Kolli:2011ki, Jang:2011vc, Kleinekathofer:2004tx, Kreisbeck:2011dh}. The bath kernel is heavily dependent on our spectral density, causing it to play a central role in our dynamics. Therefore, a more physical bath picture should provide more physically intuitive dynamics.

In our atomistic bath model, molecular vibrations in the environment (e.g. a protein framework or solvation effects) create fluctuations in the energy gaps between the ground and excited states of the system (e.g. a set of chromophores).  These time-dependent energy gaps are computed from TDDFT calculations run on each of the chromophores at each step of the MD simulation.  The key object in the computation of spectral densities is the correlation function of the energy gap time series,
\begin{equation}
C(t) = Tr_b[\hat{\Delta}(t) \hat{\Delta}(0) \hat{\rho}_b],
\end{equation}
where $\hat{\Delta}(t)$ is the time-dependent energy gap between the ground and the first excited state of the system (as calculated with TDDFT), $\hat{\rho}_b$ is the density matrix of the bath at thermal equilibrium, and $C(t)$ is the correlation function obtained after tracing over all the modes of the bath. We discretize this equation by using an unbiased autocorrelation function,
\begin{equation}\label{eq:CT}
C_k = \frac{1}{N-k} \sum_{i=1}^{N-k} (\Delta_i - \bar{\Delta}) (\Delta_{i + k} - \bar{\Delta}),
\end{equation}
where $\bar{\Delta}$ is the mean energy gap and $i$ and $k$ denote discrete time indices.  Note that $C_k$ involves comparing energy gaps that are $k$ time steps apart ($\Delta_i$ and $\Delta_{i+k}$), and $N-k$ is the total number of included comparisons.

The frequency-dependent spectral density, $J(\omega)$, is typically obtained by computing the Fourier transform of the correlation function~\cite{Valleau:2012ig}. From the definition of $C_k$ above, it is easy to check that the correlation function is real and symmetric ($C_k$ = $C_{N-k}$), which implies that the Fourier transform should be real and symmetric as well.  Because quantum mechanical spectral densities must instead be antisymmetric and obey detailed balance, it is necessary to introduce a prefactor that enforces these two properties.  Many choices are possible~\cite{Berens:1981uy}, but Valleau {\it et al.} have previously shown that a harmonic prefactor, $\beta \hbar \omega/2$, produces the most physical temperature dependence~\cite{Valleau:2012ig}.  With this choice, the spectral density becomes the cosine transform
\begin{equation}\label{eq:JCT}
J(\omega) = \frac{\beta \hbar \omega}{2} \int^{\infty}_{-\infty} \cos(\omega t) C(t) \textrm{dt}, 
\end{equation}
which characterizes the frequency-dependent coupling strength of the system to all of the nuclear vibrational modes. 

The standard approach to performing this integral is the fast Fourier transform.  Unfortunately, the fast Fourier transform requires sampling on a uniform grid at the Shannon sampling rate. This means that a relatively long time series, $C(t)$, must be computed in order to obtain good resolution of the spectral density in the frequency domain~\cite{Candes:eq, Mallat:2008vn}.  Given the computational cost of MD simulations, and the even greater expense of running TDDFT calculations on top of these simulations, any method which can reduce the required length of the time series $C(t)$ unplugs the computational bottleneck in deriving physically-accurate atomistic spectral densities.  That is our main goal in this paper.

While reducing the amount of time required to reproduce $J(\omega)$ we also choose a basis of functions which has a convenient physical form. When decomposed into a basis of damped cosines,
\begin{equation}
g_{ij}(t) = e^{-\gamma_i \, t} \cos(\Omega_j t),
\end{equation}
the function $C(t)$ is smooth and sparse.  This allows for the use of the machinery of super-resolution. 

To apply the super-resolution method, we discretize in time and cast our task as an inversion problem
\begin{equation}\label{eq:time}
C_k = \lambda_{ij} e^{-\gamma_i \, t} \cos(\Omega_j t_k),
\end{equation}
where we seek the basis expansion coefficients $\lambda_{ij}$ and have assumed Einstein summation convention over repeated indices. This can be rewritten as
\begin{equation}\label{eq:inversion}
C_k = A_{ijk} \lambda_{ij},
\end{equation}
where
\begin{equation}\label{eq:matrix}
A_{ijk} = e^{-\gamma_i \, t} cos(\Omega_j t_k)
\end{equation}
is a matrix of damped cosines, and $\lambda_{ij}$ is the set of basis coefficients we seek to recover.

The central idea of super-resolution is that the sparsity of $\lambda_{ij}$ enables its full recovery even when the system $C_k = A_{ijk} \lambda_{ij}$ is underdetermined, which is to say the number of time samples $C_k$ is significantly smaller than the number of total expansion coefficients $\lambda_{ij}$ we seek to recover.  Hence, we can recover the expansion coefficients on a dense grid of frequencies $\Omega_j$ and damping coefficients $\gamma_i$ from fewer time samples $C_k$.  Of the many possible solutions to our underdetermined system, super-resolution simply selects a balance between the smoothest and sparsest (with an emphasis on smoothness) set of basis expansion coefficients.  Formally, this is done by finding the vector $\lambda_{ij}$ that minimizes 
\begin{equation}\label{eq:CS}
\begin{aligned}
\underset{\lambda_{ij}}{\text{argmin}} \left \{ ||\nabla \lambda_{ij} ||_1 + \mu ||\lambda_{ij} ||_1 \right \}  \\
\textrm{subject to} \;\;  ||A_{ijk} \lambda_{ij} - C_k||_2 < \eta,
\end{aligned}
\end{equation}
where the subscript 1 represents the $\mathcal{L}_1$ norm (sum of absolute values), $\mu$ represents a sparsity penalty, $\nabla \lambda_{ij}$ represents the total variation norm, and $\eta$ represents the solution tolerance.  By minimizing $ ||\nabla \lambda_{ij} ||_1$, or total variation term, we are enforcing smoothness in the time domain on the reconstructed signal. This throws out the ``peaky'' solutions that can appear with compressed sensing~\cite{Andrade28082012,Sanders:2012tk}. The total variation norm also provides us with a provably exact technique for recovering peak position at the expense of peak amplitude~\cite{CPA:CPA21455}, which solves one of the issues seen previously with compressed sensing~\cite{Andrade28082012}.

Recovering the expansion coefficients $\lambda_{ij}$ in this manner by solving an underdetermined matrix inversion problem takes advantage of the natural sparsity of the problem and, as we will see in the next section, enables the construction of a well-resolved spectral density with far less time-domain data.  Even more attractive, with the $\lambda_{ij}$ coefficients in hand, it is possible to construct an analytical representation of the spectral density by taking the cosine transform of the basis functions $g_{ij}(t)$ and applying the appropriate prefactors:
\begin{equation}
\begin{aligned}
J(\omega) =  \frac{\lambda_{ij}}{\sqrt{\pi}} \left( \frac{\beta \hbar \omega \gamma_i }{\gamma_i^2 + \left( \omega - \Omega_j \right)^2 } + \frac{ \beta \hbar \omega \gamma_i }{\gamma_i^2 + \left( \omega + \Omega_j \right)^2 }  \right ),
\end{aligned}
\end{equation}
where the Einstein summation convention has again been assumed. This is an analytical representation of the spectral density in Drude-Lorentz form, and it explicitly provides the oscillation frequencies which characterize the system-bath coupling. We note that the Drude-Lorentz basis naturally provides us with a width parameter, $\gamma$, that can be understood as the lifetime of oscillations in the bath. This is seen by examining the time dependent formula, Eq. \eqref{eq:time}, where this $\gamma$ parameter determines the strength of damping. It is important to note that in the limit as $\gamma \rightarrow 0$, we recover the cosine basis in the time domain and a Dirac delta distribution in the frequency domain. By using this super-resolution technique in concert with the Drude-Lorentz basis, we see that we can recover a small set of peaks with physically-relevant information. Additionally, the parameters that characterize the Drude-Lorentz spectral densities can be input directly into both TCL-2 and HEOM without any additional parameter fitting or numerical integration.

\section{Numerical Methods}

We employ the proposed Drude-Lorentz super-resolution method described above and apply it to a monomer of the Fenna-Matthews-Olsen (FMO) photosynthetic energy transfer complex of the green-sulfur bacterium {\it C. tepidium}.  The FMO monomer is a system of seven chlorophyll molecules which are excitonically coupled to each other, as well as to the vibrations of the atoms in the protein framework.  It functions as a molecular excitonic wire, passing excitons from the light harvesting antenna complex to the reaction center, where a biochemical cascade is initiated.  

To create spectral densities for the FMO complex, we use the MD-TDDFT results of Shim {\it et al.}~\cite{Shim2012649}. The calculations were done in an isothermal-isobaric ensemble at 77 K using the AMBER force field~\cite{Cornell:1995td,Ceccarelli:2003ts}. These calculations began with a 2 ns equilibration before performing the production computations. The production steps ran for a total of 40 picoseconds with a 2 femtosecond timestep, and the optical gap was calculated for each fragment every 4 femtoseconds using TDDFT with the BLYP~\cite{Becke:1988tx, Miehlich:1989vq, Lee:1988ub} functional in the 3-21G basis set in Q-Chem~\cite{Shao:2006wf}. 

To perform super-resolution numerically, we require an algorithm which minimizes the total variation norm to solve the minimization problem described by eq.~\eqref{eq:CS}.  In our implementation, we use the two step iterative shrinkage thresholding (TwIST) algorithm~\cite{BioucasDias:2007tp, 4378902}, which combines computational efficiency with strong convergence.  To construct the measurement matrix $A$ described in eq.~\eqref{eq:matrix}, we must select a grid of possible frequencies ($\left\{\Omega_j\right\}$) and linewidths ($\left\{\gamma_i\right\}$).  In our implementation, we use a grid of frequencies ranging from 0 to 2000 cm$^{-1}$ in 2 cm$^{-1}$ intervals, and a grid of linewidths ranging from 0 to 160 cm$^{-1}$ in 6 cm$^{-1}$ intervals.  We assume that our calculations are converged when $\eta < 10^{-7}$ (in eq.~\eqref{eq:CS}), or the solution vector remains constant for 100 iterations.  Finally, we perform an $\mathcal{L}_2$ minimization of $A_{ijk} \lambda_{ij} - C_k$ while freezing the recovered nonzero basis functions, allowing us to further minimize the error. We refer to this procedure as debiasing because it partly removes the bias towards sparsity and smoothness introduced by the $\mathcal{L}_1$ minimization. This debiasing procedure reduces our solution tolerance to $\eta < 10^{-9}$, allowing convergence to a better solution. It is important to note that, in general, the super-resolution technique is robust to an over-complete basis.

\section{Results}
\begin{figure}
\begin{center}
\includegraphics[width=\columnwidth]{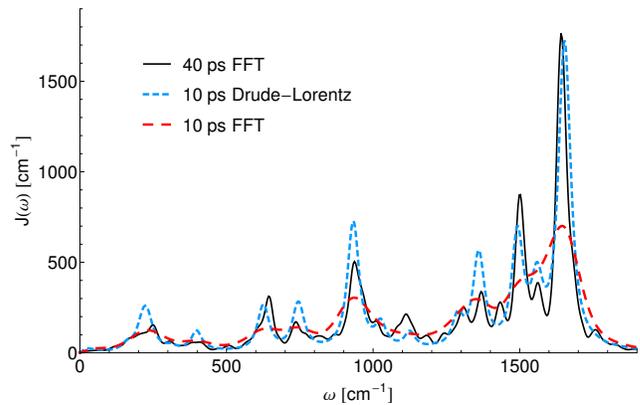}
\caption{ Comparison of the spectral density for site 1 of the FMO complex as a function of time and technique for spectral density recovery. Compared to the fast Fourier transform at 40 ps, much of the fine structure is easily recovered by super-resolution in the Drude-Lorentz basis, even with significant undersampling by a factor of four. }
\label{fig:sds}
\end{center}
\end{figure}

Fig.~\ref{fig:sds} shows the results of employing the Drude-Lorentz super-resolution method to recover the spectral density for site 1 of FMO.  The figure compares Drude-Lorentz super-resolution with 10 ps of MD to a standard fast Fourier transform approach with both 10 and 40 ps of MD. We take the fast Fourier transform with 40 ps of MD as our standard for comparison.  By comparing the two methods with 10 ps of MD, it is clear that super-resolution resolves more features of the spectral density than the standard fast Fourier transform from the same amount of time-domain data.  Moreover, super-resolution captures most of the features of the fast Fourier transform with the full 40 ps of MD: we see the expected CO stretch at 1600 cm$^{-1}$, which we attribute to the amides in the protein scaffold, as well as all of the other major peaks in the spectral density. We attribute a significant amount of the error in our spectral density reconstruction to the fact that the truncated MD series does not explore the phase space as thoroughly in only 10 ps. 

The Drude-Lorentz basis also provides significant sparsity gains in comparison to the cosine basis: we require only 56 Drude-Lorentz peaks to create the spectral density given in Fig.~\ref{fig:sds} whereas. This sparsity provides a significant computational advantage for excitonic propagation in both hierarchical equations of motion (HEOM)~\cite{JPSJ.58.101} and second order time-convolutionless master equation~\cite{Breuer:2007wp}(TCL-2) approaches because the propagations scale factorially and linearly, respectively, as a function of the number of peaks included. In the excitonically accessible regime of 0-540cm$^{-1}$, we recover only 20 Drude-Lorentz peaks, and six of them have amplitudes that are two orders of magnitude smaller than the rest. These Drude-Lorentz peaks can be entered directly into master equation simulations, including HEOM codes, without the need to perform any intermediate fitting~\cite{16106}. In summary, super-resolution yields a well-resolved spectral density using less time-domain data than is required by the standard fast Fourier transform approach and precludes the need for additional fitting.

As mentioned above, the TCL-2 propagation of the exciton dynamics of the FMO complex, with the Hamiltonian coming from~\cite{Adolphs:2006ey}, was carried out using the Drude-Lorenz spectral densities obtained from super-resolution. We propagated 1 ps of dynamics and obtained the populations of sites~1-3, as well as the coherence between sites 1 and 3.

\begin{figure}
\begin{center}
\includegraphics[width=\columnwidth]{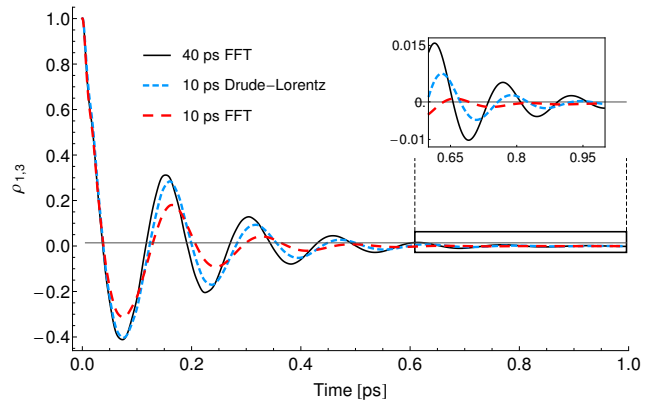}
\caption{ Comparison of the coherences between excitonic eigenstates 1 and 3 as a function of time and technique for spectral density recovery. Compared to the fast Fourier transform at 40 ps, the 10 ps Drude-Lorentz decomposition introduces a slight shift in oscillation frequency, but nevertheless yields more accurate dynamics than the equivalently-sampled fast Fourier transform at 10 ps. }
\label{fig:coh}
\end{center}
\end{figure}

Fig.~\ref{fig:coh} shows the coherence between excitonic eigenstates 1 and 3 as a function of time.  Compared to the 40 ps fast Fourier transform, we see that the 10 ps Drude-Lorentz super-resolution more faithfully reproduces the coherence dynamics than the 10 ps fast Fourier transform, both in terms of the oscillation frequency and the overall damping.  The fast Fourier transform with 10 ps of MD data introduces serious overdamping as well as a significant shift in oscillation frequency.  In contrast, the Drude-Lorentz expansion with 10 ps of MD data introduces only a small shift in oscillation frequency, resulting in more accurate coherence dynamics overall.  We attribute most of the discrepancies to slight relative differences in the reorganization of each site between spectral densities constructed with 10 and 40 ps of MD data. It appears that while the oscillations are extremely sensitive to the relative reorganization energies between the sites, the damping is more dependent on the fine structure of the spectral densities. The Drude-Lorentz super-resolution (10 ps MD-TDDFT data) reproduces the coherence life-times obtained by fast Fourier transform recovered using all 40 ps of MD-TDDFT data -- representing a factor of four improvement.

\begin{figure}
\begin{center}
\includegraphics[width=\columnwidth]{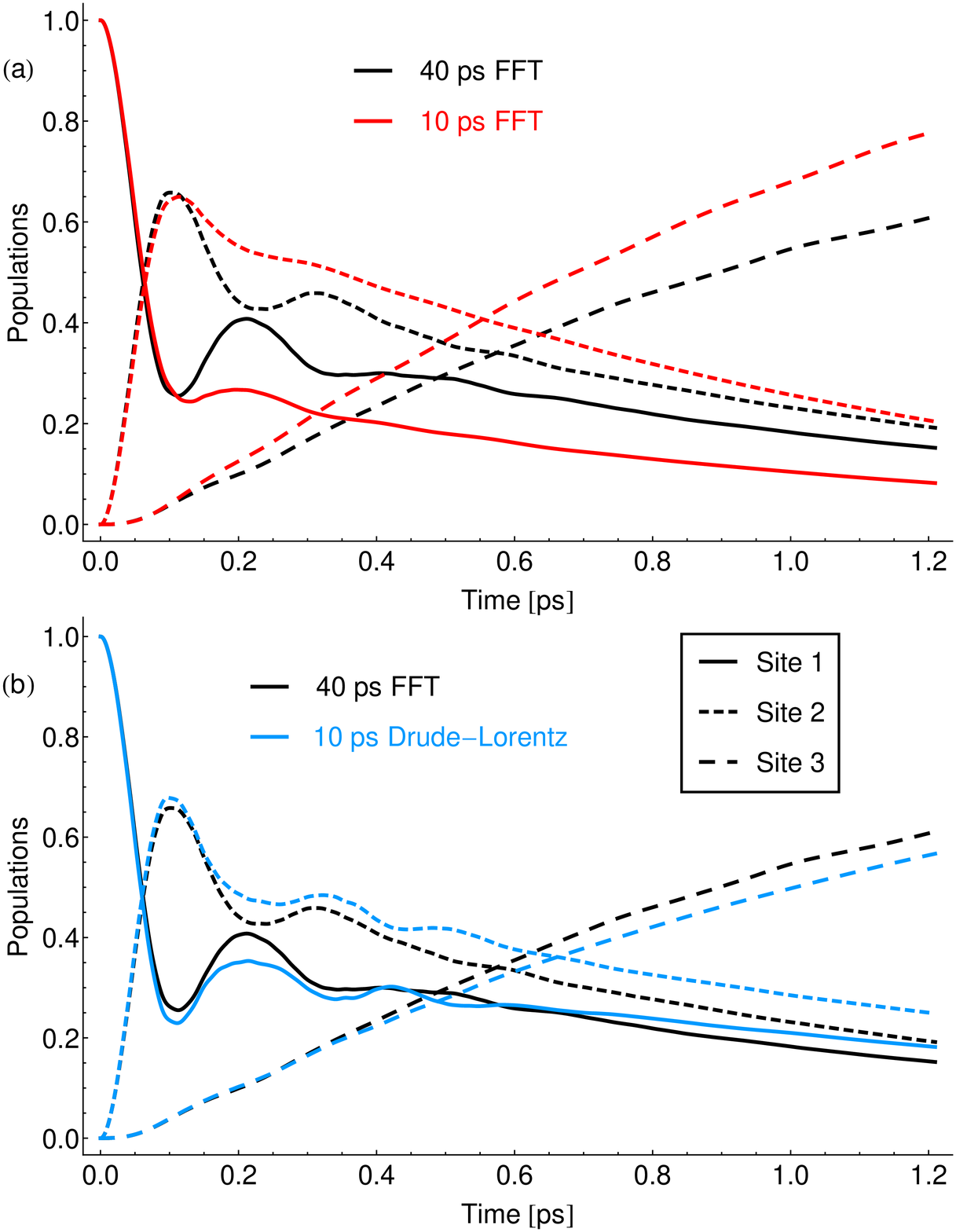}
\caption{ Comparison of the populations for sites 1-3 as a function of time and technique for spectral density recovery Compared to the fast Fourier transform at 40 ps, the 10 ps Drude-Lorentz decomposition recovers the overall shape and provides much more faithful dynamics than the equivalently-sampled fast Fourier transform at 10 ps. }
\label{fig:pop}
\end{center}
\end{figure}

The contrast between the two approximation techniques becomes even more significant when we simulate dynamics beginning with an exciton fully localized on site 1. In Fig.~\ref{fig:pop}, we have plotted the populations of the first three sites as a function of time. The Drude-Lorentz expansion with 10 ps of MD yields good qualitative agreement with our standard of comparison whereas the fast Fourier transform. The fast Fourier transform on 10 ps overestimates population transfer to site 3 at short times and grows much more quickly from there, whereas the Drude-Lorentz expansion slightly under predicts the population transfer at long times. We attribute these errors in the asymptotic behavior to slight differences in the reorganization energies for the spectral densities of each of the sites, since each site is embedded in a different enviornment, the reorganization process of the individual pigments is different. This sensitivity affects overall dissipation and even small changes in the spectral density of the Drude-Lorentz expansion (10 ps) when compared to the standard of comparison affects energy relaxation. Beyond that, the Drude-Lorentz expansion is capable of reproducing the oscillations at 0.2 and 0.4 ps in the data for sites 1 and 2 whereas the fast Fourier transform reproduces them less faithfully. In summary, the Drude-Lorentz super-resolution technique provides us with much more physical behavior.

\section{Conclusions}
We have shown that the Drude-Lorentz super-resolution method provides significant computational advantages for the construction of atomistic bath models.  In particular, the super-resolution calculations require only 10 ps of MD-TDDFT simulations to obtain reasonable atomistic spectral densities and system dynamics; this is one quarter the amount of data needed in standard fast Fourier transform-based calculations. Ultimately, this will permit the use of more physically accurate calculations or larger systems. Given the computational expense of running TDDFT calculations at every MD simulation step, we believe that the super-resolution method will enable the treatment of larger systems than previously possible. 

One of the most significant advantages of our super-resolution method is the decomposition of these atomistic spectral densities into a naturally-sparse basis of Drude-Lorentz oscillators.  This makes it easy to perform fast master equation simulations within either the TCL-2 or HEOM formalisms by exploiting analytic integrals of the spectral density.  Beyond this, we also directly extract physically-important parameters such as the coherence lifetimes of all the oscillators in the bath.  In the future, it is easy to imagine turning this technique on its head to create new spectral densities in a constructive fashion from a set of Drude-Lorentz oscillators.

\section{Acknowledgements} 
We acknowledge S. Valleau for useful discussions and computer code. We acknowledge the financial support of Defense Advanced Research Projects Agency grant N66001-10-1-4063 and the Defense Threat Reduction Agency under contract no. HDTRA1-10-1-0046. T.M. acknowledges support from the National Science Foundation (NSF) through the Graduate Research Fellowship Program (GRFP). S.B. acknowledges support from the Department of Energy (DoE) through the Computational Sciences Graduate Fellowship (CSGF). J.N.S. acknowledges support from the Department of Defense (DoD) through the National Defense Science {\&} Engineering Graduate Fellowship (NDSEG) Program. A.A.G. thanks the Corning Foundation.


\begin{thebibliography}{10}

\bibitem{Breuer:2007wp}
{\sc H.-P. Breuer} and {\sc F.~Petruccione},
\newblock {\em {The Theory of Open Quantum Systems}},
\newblock Oxford University Press, USA, 2007.

\bibitem{Shenvi:2008tr}
{\sc N.~Shenvi}, {\sc J.~R. Schmidt}, {\sc S.~T. Edwards}, and {\sc J.~C.
  Tully},
\newblock {\em Phys. Rev. A} {\bf 78}, 022502 (2008).

\bibitem{2013JChPh.138k4102B}
{\sc T.~C. Berkelbach}, {\sc M.~S. Hybertsen}, and {\sc D.~R. Reichman},
\newblock {\em Journal of Chemical Physics} {\bf 138}, 4102 (2013).

\bibitem{Berkelbach:2013vw}
{\sc T.~C. Berkelbach}, {\sc M.~S. Hybertsen}, and {\sc D.~R. Reichman},
\newblock {\em J. Chem. Phys.} {\bf 138}, 114103 (2013).

\bibitem{Singh:2011fl}
{\sc N.~Singh} and {\sc P.~Brumer},
\newblock {\em Faraday Discuss.} {\bf 153}, 41 (2011).

\bibitem{Jang:2007cz}
{\sc S.~Jang}, {\sc M.~D. Newton}, and {\sc R.~J. Silbey},
\newblock {\em J. Phys. Chem. B} {\bf 111}, 6807 (2007).

\bibitem{Mohseni:2008gp}
{\sc M.~Mohseni}, {\sc P.~Rebentrost}, {\sc S.~Lloyd}, and {\sc
  A.~Aspuru-Guzik},
\newblock {\em J. Chem. Phys.} {\bf 129}, 174106 (2008).

\bibitem{Plenio:2008ff}
{\sc M.~B. Plenio} and {\sc S.~F. Huelga},
\newblock {\em New J. Phys.} {\bf 10}, 113019 (2008).

\bibitem{Rebentrost:2009hu}
{\sc P.~Rebentrost}, {\sc M.~Mohseni}, {\sc I.~Kassal}, {\sc S.~Lloyd}, and
  {\sc A.~Aspuru-Guzik},
\newblock {\em New J. Phys.} {\bf 11}, 033003 (2009).

\bibitem{Cao:2009cc}
{\sc J.~Cao} and {\sc R.~J. Silbey},
\newblock {\em J. Phys. Chem. A} {\bf 113}, 13825 (2009).

\bibitem{Ishizaki:2009tt}
{\sc A.~Ishizaki} and {\sc G.~R. Fleming},
\newblock {\em Proc. Natl. Acad. Sci.} {\bf 106}, 17255 (2009).

\bibitem{Ishizaki:2011cx}
{\sc A.~Ishizaki} and {\sc G.~R. Fleming},
\newblock {\em J. Phys. Chem. B} {\bf 115}, 6227 (2011).

\bibitem{Sarovar:2010hs}
{\sc M.~Sarovar}, {\sc A.~Ishizaki}, {\sc G.~R. Fleming}, and {\sc K.~B.
  Whaley},
\newblock {\em Nature Phys.} {\bf 6}, 462 (2010).

\bibitem{Abramavicius:2010et}
{\sc D.~Abramavicius} and {\sc S.~Mukamel},
\newblock {\em J. Chem. Phys.} {\bf 133}, 064510 (2010).

\bibitem{Wu:2010bg}
{\sc J.~Wu}, {\sc F.~Liu}, {\sc Y.~Shen}, {\sc J.~Cao}, and {\sc R.~J. Silbey},
\newblock {\em New J. Phys.} {\bf 12}, 105012 (2010).

\bibitem{Moix:2011gi}
{\sc J.~Moix}, {\sc J.~Wu}, {\sc P.~Huo}, {\sc D.~Coker}, and {\sc J.~Cao},
\newblock {\em J. Phys. Chem. Lett.} {\bf 2}, 3045 (2011).

\bibitem{Kreisbeck:2011dh}
{\sc C.~Kreisbeck}, {\sc T.~Kramer}, {\sc M.~Rodr{\'\i}guez}, and {\sc
  B.~Hein},
\newblock {\em J. Chem. Theory Comput.} {\bf 7}, 2166 (2011).

\bibitem{Skochdopole:2011gh}
{\sc N.~Skochdopole} and {\sc D.~A. Mazziotti},
\newblock {\em J. Phys. Chem. Lett.} {\bf 2}, 2989 (2011).

\bibitem{Ritschel:2011ht}
{\sc G.~Ritschel}, {\sc J.~Roden}, {\sc W.~T. Strunz}, {\sc A.~Aspuru-Guzik},
  and {\sc A.~Eisfeld},
\newblock {\em J. Phys. Chem. Lett.} {\bf 2}, 2912 (2011).

\bibitem{Rebentrost:2011hc}
{\sc P.~Rebentrost} and {\sc A.~Aspuru-Guzik},
\newblock {\em J. Chem. Phys.} {\bf 134}, 101103 (2011).

\bibitem{Pachon:2012fm}
{\sc L.~A. Pach{\'o}n} and {\sc P.~Brumer},
\newblock {\em Phys. Chem. Chem. Phys.} {\bf 14}, 10094 (2012).

\bibitem{Vlaming:2012hv}
{\sc S.~M. Vlaming} and {\sc R.~J. Silbey},
\newblock {\em J. Chem. Phys.} {\bf 136}, 055102 (2012).

\bibitem{Caruso:2012gf}
{\sc F.~Caruso}, {\sc S.~K. Saikin}, {\sc E.~Solano}, {\sc S.~F. Huelga}, {\sc
  A.~Aspuru-Guzik}, and {\sc M.~B. Plenio},
\newblock {\em Phys. Rev. B} {\bf 85}, 125424 (2012).

\bibitem{Zhu:2011ea}
{\sc J.~Zhu}, {\sc S.~Kais}, {\sc P.~Rebentrost}, and {\sc A.~Aspuru-Guzik},
\newblock {\em J. Phys. Chem. B} {\bf 115}, 1531 (2011).

\bibitem{Roden:2009gr}
{\sc J.~Roden}, {\sc A.~Eisfeld}, {\sc W.~Wolff}, and {\sc W.~Strunz},
\newblock {\em Phys. Rev. Lett.} {\bf 103}, 058301 (2009).

\bibitem{Olbrich:2010ce}
{\sc C.~Olbrich} and {\sc U.~Kleinekath{\"o}fer},
\newblock {\em J. Phys. Chem. B} {\bf 114}, 12427 (2010).

\bibitem{Olbrich:2011hc}
{\sc C.~Olbrich}, {\sc T.~L.~C. Jansen}, {\sc J.~Liebers}, {\sc M.~Aghtar},
  {\sc J.~Str{\"u}mpfer}, {\sc K.~Schulten}, {\sc J.~Knoester}, and {\sc
  U.~Kleinekath{\"o}fer},
\newblock {\em J. Phys. Chem. B} {\bf 115}, 8609 (2011).

\bibitem{Hein:2012vn}
{\sc B.~Hein}, {\sc C.~Kreisbeck}, {\sc T.~Kramer}, and {\sc
  M.~Rodr{\'\i}guez},
\newblock {\em New J. Phys.} {\bf 14}, 023018 (2012).

\bibitem{doi:10.1021/jp304649c}
{\sc N.~Christensson}, {\sc H.~F. Kauffmann}, {\sc T.~Pullerits}, and {\sc
  T.~Man{\v c}al},
\newblock {\em J. Phys. Chem. B} {\bf 116}, 7449 (2012).

\bibitem{Chin:2013uh}
{\sc A.~W. Chin}, {\sc J.~Prior}, {\sc R.~Rosenbach}, {\sc F.~Caycedo-Soler},
  {\sc S.~F. Huelga}, and {\sc M.~B. Plenio},
\newblock {\em Nature Phys.} {\bf 9}, 113 (2013).

\bibitem{Kreisbeck:2012ui}
{\sc C.~Kreisbeck} and {\sc T.~Kramer},
\newblock {\em J. Phys. Chem. Lett.} {\bf 3}, 2828 (2012).

\bibitem{Valleau:2012ig}
{\sc S.~Valleau}, {\sc A.~Eisfeld}, and {\sc A.~Aspuru-Guzik},
\newblock {\em J. Chem. Phys.} {\bf 137}, 224103 (2012).

\bibitem{Tuckerman:2010wy}
{\sc M.~Tuckerman},
\newblock {\em Statistical Mechanics: Theory and Molecular Simulation},
\newblock OUP Oxford, 2010.

\bibitem{Runge:1984us}
{\sc E.~Runge} and {\sc E.~K. Gross},
\newblock {\em Phys. Rev. Lett.} {\bf 52}, 997 (1984).

\bibitem{Mallat:2008vn}
{\sc S.~Mallat},
\newblock {\em A Wavelet Tour of Signal Processing, Third Edition: The Sparse
  Way},
\newblock Academic Press, 3 edition, 2008.

\bibitem{Freeman:2002va}
{\sc W.~T. Freeman}, {\sc T.~R. Jones}, and {\sc E.~C. Pasztor},
\newblock {\em IEEE Comput. Graph. Appl.} {\bf 22}, 56 (2002).

\bibitem{Patti:1997vq}
{\sc A.~J. Patti}, {\sc M.~I. Sezan}, and {\sc A.~Murat~Tekalp},
\newblock {\em IEEE Trans. Image Process.} {\bf 6}, 1064 (1997).

\bibitem{Elad:1997un}
{\sc M.~Elad} and {\sc A.~Feuer},
\newblock {\em IEEE Trans. Image Process.} {\bf 6}, 1646 (1997).

\bibitem{2005A&A...436..373P}
{\sc K.~G. Puschmann} and {\sc F.~Kneer},
\newblock {\em Astron. Astrophysic.} {\bf 436}, 373 (2005).

\bibitem{Mccutchen:1967uf}
{\sc C.~W. Mccutchen},
\newblock {\em J. Opt. Soc. Am.} {\bf 57}, 1190 (1967).

\bibitem{5193030}
{\sc D.~Kouame} and {\sc M.~Ploquin},
\newblock {Super-resolution in medical imaging: An illustrative approach
  through ultrasound},
\newblock in {\em Biomedical Imaging: From Nano to Macro, 2009. ISBI '09. IEEE
  International Symposium on}, pp. 249--252, 2009.

\bibitem{5272200}
{\sc J.~Ma},
\newblock {\em Instrumentation and Measurement, IEEE Transactions on} {\bf 59},
  1600 (2010).

\bibitem{Donoho:ci}
{\sc D.~L. Donoho},
\newblock {\em IEEE Trans. Inform. Theory} {\bf 52}, 1289 (2006).

\bibitem{5288845}
{\sc A.~Oka} and {\sc L.~Lampe},
\newblock {A compressed sensing receiver for bursty communication with UWB
  Impulse Radio},
\newblock in {\em Ultra-Wideband, 2009. ICUWB 2009. IEEE International
  Conference on}, pp. 279--284, 2009.

\bibitem{4770164}
{\sc M.~A. Herman} and {\sc T.~Strohmer},
\newblock {\em Signal Processing, IEEE Transactions on} {\bf 57}, 2275 (2009).

\bibitem{4959603}
{\sc C.~Qiu}, {\sc W.~Lu}, and {\sc N.~Vaswani},
\newblock {Real-time dynamic MR image reconstruction using Kalman Filtered
  Compressed Sensing},
\newblock in {\em Acoustics, Speech and Signal Processing, 2009. ICASSP 2009.
  IEEE International Conference on}, pp. 393--396, 2009.

\bibitem{MRM:MRM21391}
{\sc M.~Lustig}, {\sc D.~Donoho}, and {\sc J.~M. Pauly},
\newblock {\em Magnetic Resonance in Medicine} {\bf 58}, 1182 (2007).

\bibitem{6426647}
{\sc M.~Nagahara}, {\sc D.~E. Quevedo}, and {\sc J.~Ostergaard},
\newblock {Sparse representations for packetized predictive networked control},
\newblock in {\em Decision and Control (CDC), 2012 IEEE 51st Annual Conference
  on}, pp. 1362--1367, 2012.

\bibitem{Tuma:2009gb}
{\sc T.~Tuma}, {\sc S.~Rooney}, and {\sc P.~Hurley},
\newblock {On the Applicability of Compressive Sampling in Fine Grained
  Processor Performance Monitoring},
\newblock in {\em 2009 14th IEEE International Conference on Engineering of
  Complex Computer Systems}, pp. 210--219, IEEE, 2009.

\bibitem{Shabani:2009de}
{\sc A.~Shabani}, {\sc R.~L. Kosut}, {\sc M.~Mohseni}, {\sc H.~Rabitz}, {\sc
  M.~A. Broome}, {\sc M.~P. Almeida}, {\sc A.~Fedrizzi}, and {\sc A.~G. White},
\newblock {\em arXiv}  (2009).

\bibitem{5419072}
{\sc M.~Mishali} and {\sc Y.~C. Eldar},
\newblock {\em Selected Topics in Signal Processing, IEEE Journal of} {\bf 4},
  375 (2010).

\bibitem{4472247}
{\sc M.~F. Duarte}, {\sc M.~A. Davenport}, {\sc D.~Takhar}, {\sc J.~N. Laska},
  {\sc T.~Sun}, {\sc K.~F. Kelly}, and {\sc R.~G. Baraniuk},
\newblock {\em Signal Processing Magazine, IEEE} {\bf 25}, 83 (2008).

\bibitem{Coulter:2010wx}
{\sc W.~K. Coulter}, {\sc C.~J. Hillar}, {\sc G.~Isley}, and {\sc F.~T.
  Sommer},
\newblock p. 5494 (2010).

\bibitem{CPA:CPA21455}
{\sc E.~J. Cand{\`e}s} and {\sc C.~Fernandez-Granda},
\newblock {\em Comm. Pure Appl. Math.}  (2013).

\bibitem{BioucasDias:2007tp}
{\sc J.~M. Bioucas-Dias} and {\sc M.~A. Figueiredo},
\newblock {\em IEEE Trans. Image Process.} {\bf 16}, 2992 (2007).

\bibitem{4378902}
{\sc J.~M. Bioucas-Dias} and {\sc M.~A.~T. Figueiredo},
\newblock {Two-step algorithms for linear inverse problems with non-quadratic
  regularization},
\newblock in {\em Image Processing, 2007. ICIP 2007. IEEE International
  Conference on}, pp. I--105--I -- 108, 2007.

\bibitem{Vulto:1998bu}
{\sc S.~I.~E. Vulto}, {\sc M.~A. de~Baat}, {\sc R.~J.~W. Louwe}, {\sc H.~P.
  Permentier}, {\sc T.~Neef}, {\sc M.~Miller}, {\sc H.~van Amerongen}, and {\sc
  T.~J. Aartsma},
\newblock {\em J. Phys. Chem. B} {\bf 102}, 9577 (1998).

\bibitem{Adolphs:2006ey}
{\sc J.~Adolphs} and {\sc T.~Renger},
\newblock {\em Biophys. J.} {\bf 91}, 2778 (2006).

\bibitem{Mohseni:2011wg}
{\sc M.~Mohseni}, {\sc A.~Shabani}, {\sc S.~Lloyd}, and {\sc H.~Rabitz},
\newblock {\em arXiv}  (2011).

\bibitem{Chen:2013wz}
{\sc X.~Chen}, {\sc J.~Cao}, and {\sc R.~J. Silbey},
\newblock {\em J. Chem. Phys.}  (2013).

\bibitem{YuenZhou:2012vf}
{\sc J.~Yuen-Zhou}, {\sc J.~J. Krich}, and {\sc A.~Aspuru-Guzik},
\newblock {\em J. Chem. Phys.} {\bf 136}, 234501 (2012).

\bibitem{Shim2012649}
{\sc S.~Shim}, {\sc P.~Rebentrost}, {\sc S.~p. Valleau}, and {\sc
  A.~n~Aspuru-Guzik},
\newblock {\em Biophys. J.} {\bf 102}, 649 (2012).

\bibitem{Kolli:2011vy}
{\sc A.~Kolli}, {\sc A.~Nazir}, and {\sc A.~Olaya-Castro},
\newblock {\em J. Chem. Phys.} {\bf 135}, 154112 (2011).

\bibitem{Jang:2011vc}
{\sc S.~Jang},
\newblock {\em J. Chem. Phys.} {\bf 135}, 034105 (2011).

\bibitem{Ishizaki:2009uh}
{\sc A.~Ishizaki} and {\sc G.~R. Fleming},
\newblock {\em J. Chem. Phys.} {\bf 130}, 234111 (2009).

\bibitem{JPSJ.58.101}
{\sc Y.~Tanimura} and {\sc R.~Kubo},
\newblock {\em Journal of the Physical Society of Japan} {\bf 58}, 101 (1989).

\bibitem{16106}
{\sc C.~Kreisbeck} and {\sc T.~Kramer},
\newblock Exciton Dynamics Lab for Light-Harvesting Complexes (GPU-HEOM), 2013.

\bibitem{Rebentrost:2011vh}
{\sc P.~Rebentrost} and {\sc A.~Aspuru-Guzik},
\newblock {\em J. Chem. Phys.} {\bf 134}, 101103 (2011).

\bibitem{2006JChPh.125j4906P}
{\sc A.~Pereverzev} and {\sc E.~R. Bittner},
\newblock {\em Journal of Chemical Physics} {\bf 125}, 4906 (2006).

\bibitem{2011PhRvB..83k5416T}
{\sc C.~Timm},
\newblock {\em Phys. Rev. B} {\bf 83}, 115416 (2011).

\bibitem{Ahn:1994ww}
{\sc D.~Ahn},
\newblock {\em Phys. Rev. B} {\bf 50}, 8310 (1994).

\bibitem{HeinzPeter:2000ul}
{\sc B.~Heinz-Peter}, {\sc B.~Kappler}, and {\sc F.~Petruccione},
\newblock {\em Decoherence: Theoretical, Experimental, and Conceptual Problems}
  , 233 (2000).

\bibitem{Shabani:2005wa}
{\sc A.~Shabani} and {\sc D.~A. Lidar},
\newblock {\em Phys. Rev. A} {\bf 71}, 020101 (2005).

\bibitem{Smirne:2010te}
{\sc A.~Smirne} and {\sc B.~Vacchini},
\newblock {\em Phys. Rev. A} {\bf 82}, 022110 (2010).

\bibitem{Kolli:2011ki}
{\sc A.~Kolli}, {\sc A.~Nazir}, and {\sc A.~Olaya-Castro},
\newblock {\em J. Chem. Phys.} {\bf 135}, 154112 (2011).

\bibitem{Kleinekathofer:2004tx}
{\sc U.~Kleinekath{\"o}fer},
\newblock {\em J. Chem. Phys.} {\bf 121}, 2505 (2004).

\bibitem{Berens:1981uy}
{\sc P.~H. Berens}, {\sc S.~R. White}, and {\sc K.~R. Wilson},
\newblock {\em J. Chem. Phys.} {\bf 75}, 515 (1981).

\bibitem{Candes:eq}
{\sc E.~J. Candes}, {\sc J.~Romberg}, and {\sc T.~Tao},
\newblock {\em IEEE Trans. Inform. Theory} {\bf 52}, 489 (2006).

\bibitem{Andrade28082012}
{\sc X.~Andrade}, {\sc J.~N. Sanders}, and {\sc A.~Aspuru-Guzik},
\newblock {\em Proc. Natl. Acad. Sci.} {\bf 109}, 13928 (2012).

\bibitem{Sanders:2012tk}
{\sc J.~N. Sanders}, {\sc S.~K. Saikin}, {\sc S.~Mostame}, {\sc X.~Andrade},
  {\sc J.~R. Widom}, {\sc A.~H. Marcus}, and {\sc A.~Aspuru-Guzik},
\newblock {\em J. Phys. Chem. Lett.} {\bf 3}, 2697 (2012).

\bibitem{Cornell:1995td}
{\sc W.~D. Cornell}, {\sc P.~Cieplak}, {\sc C.~I. Bayly}, {\sc I.~R. Gould},
  {\sc K.~M. Merz}, {\sc D.~M. Ferguson}, {\sc D.~C. Spellmeyer}, {\sc T.~Fox},
  {\sc J.~W. Caldwell}, and {\sc P.~A. Kollman},
\newblock {\em J. Am. Chem. Soc.} {\bf 117}, 5179 (1995).

\bibitem{Ceccarelli:2003ts}
{\sc M.~Ceccarelli}, {\sc P.~Procacci}, and {\sc M.~Marchi},
\newblock {\em J. Comput. Chem.} {\bf 24}, 129 (2003).

\bibitem{Becke:1988tx}
{\sc A.~D. Becke},
\newblock {\em Phys. Rev. A} {\bf 38}, 3098 (1988).

\bibitem{Miehlich:1989vq}
{\sc B.~Miehlich}, {\sc A.~Savin}, {\sc H.~Stoll}, and {\sc H.~Preuss},
\newblock {\em Chem. Phys. Lett.} {\bf 157}, 200 (1989).

\bibitem{Lee:1988ub}
{\sc C.~Lee}, {\sc W.~Yang}, and {\sc R.~G. Parr},
\newblock {\em Phys. Rev. B} {\bf 37}, 785 (1988).

\bibitem{Shao:2006wf}
{\sc Y.~Shao}, {\sc L.~F. Molnar}, {\sc Y.~Jung}, {\sc J.~Kussmann}, {\sc
  C.~Ochsenfeld}, {\sc S.~T. Brown}, {\sc A.~T. Gilbert}, {\sc L.~V.
  Slipchenko}, {\sc S.~V. Levchenko}, and {\sc D.~P. O'Neill},
\newblock {\em Phys. Chem. Chem. Phys.} {\bf 8}, 3172 (2006).

\end{thebibliography}
\end{document}